\documentclass[twocolumn]{aastex701}

\usepackage{CJK}
\usepackage{xspace}
\usepackage{hyperref}

\newcommand{\Oumuamua}{\okina Oumuamua\xspace}

\newcommand{\Nitau}{$593.7\pm14.8$\xspace}
\newcommand{\CNtau}{$841.0\pm15.4$\xspace}

\hypersetup{pdfauthor={Name}}
\begin{document}
\begin{CJK*}{UTF8}{gbsn}

\DeclareRobustCommand{\okina}{%
  \raisebox{\dimexpr\fontcharht\font`A-\height}{%
    \scalebox{0.8}{`}%
  }%
}

\title{Spatial Profiles of 3I/ATLAS CN and Ni Outgassing from Keck/KCWI Integral Field Spectroscopy}

\author[0000-0003-3953-9532]{Willem~B.~Hoogendam}
\altaffiliation{NSF Graduate Research Fellow}
\affiliation{Institute for Astronomy, University of Hawai\okina i, 
2680 Woodlawn Drive, Honolulu, HI 96822, USA}
\email{willemh@hawaii.edu}  

\author[0000-0003-4631-1149]{Benjamin~J.~Shappee}
\affiliation{Institute for Astronomy, University of Hawai\okina i, 
2680 Woodlawn Drive, Honolulu, HI 96822, USA}
\email{shappee@hawaii.edu} 

\author[0000-0001-5559-2179]{James~J.~Wray}
\affiliation{School of Earth and Atmospheric Sciences, Georgia Institute of Technology, 311 Ferst Drive, Atlanta, GA 30332, USA}
\affiliation{Institute for Astronomy, University of Hawai\okina i, 
2680 Woodlawn Drive, Honolulu, HI 96822, USA}
\email{jwray@gatech.edu}

\author[0000-0002-5033-9593]{Bin Yang}
\affiliation{Instituto de Estudios Astrof\'isicos, Facultad de Ingenier\'ia y Ciencias, Universidad Diego Portales, Santiago, Chile}
\email{bin.yang@mail.udp.cl} 

\author[0000-0002-2058-5670]{Karen~J.~Meech}
\affiliation{Institute for Astronomy, University of Hawai\okina i, 
2680 Woodlawn Drive, Honolulu, HI 96822, USA}
\email{meech@hawaii.edu}  

\author[0000-0002-5221-7557]{Christopher~Ashall}
\affiliation{Institute for Astronomy, University of Hawai\okina i, 
2680 Woodlawn Drive, Honolulu, HI 96822, USA}
\email{cashall@hawaii.edu}

\author[0000-0002-2164-859X]{Dhvanil~D.~Desai}
\affiliation{Institute for Astronomy, University of Hawai\okina i, 
2680 Woodlawn Drive, Honolulu, HI 96822, USA}
\email{dddesai@hawaii.edu}

\author[0009-0003-2390-2840]{Kyle~Hart}
\affiliation{Institute for Astronomy, University of Hawai\okina i, 
2680 Woodlawn Drive, Honolulu, HI 96822, USA}
\email{kylehart@hawaii.edu}  

\author[0000-0001-9668-2920]{Jason~T.~Hinkle}
\altaffiliation{NHFP Einstein Fellow}
\affiliation{Department of Astronomy, University of Illinois Urbana-Champaign, 1002 West Green Street, Urbana, IL 61801, USA}
\affiliation{NSF-Simons AI Institute for the Sky (SkAI), 172 E. Chestnut St., Chicago, IL 60611, USA}
\affiliation{Institute for Astronomy, University of Hawai\okina i, 
2680 Woodlawn Drive, Honolulu, HI 96822, USA}
\email{jhinkle6@hawaii.edu}

\author[orcid=0000-0002-8732-6980]{Andrew~Hoffman}
\affiliation{Institute for Astronomy, University of Hawai\okina i, 
2680 Woodlawn Drive, Honolulu, HI 96822, USA}
\email{amho@hawaii.edu}  

\author[orcid=0009-0008-7427-4617]{Esther~M.~Hu}
\affiliation{Institute for Astronomy, University of Hawai\okina i, 
2680 Woodlawn Drive, Honolulu, HI 96822, USA}
\email{ehu@hawaii.edu}  

\author[0000-0002-6230-0151]{David~O.~Jones}
\affiliation{Institute for Astronomy, University of Hawai\okina i, 640 N. A\okina ohoku Pl., Hilo, HI 96720, USA}
\email{dojones@hawaii.edu}

\author[0000-0001-7186-105X]{Kyle~Medler}
\affiliation{Institute for Astronomy, University of Hawai\okina i, 
2680 Woodlawn Drive, Honolulu, HI 96822, USA}
\email{kmedler@hawaii.edu}

\author[0000-0002-7305-8321]{Cameron~M.~Pfeffer}
\affiliation{Institute for Astronomy, University of Hawai\okina i, 
2680 Woodlawn Drive, Honolulu, HI 96822, USA}
\email{cpfeffer@hawaii.edu}

\begin{abstract}
Cometary activity from interstellar objects provides a unique window into the environs of other stellar systems. We report blue-sensitive integral field unit spectroscopy of the interstellar object 3I/ATLAS from the Keck-II-mounted Keck Cosmic Web Imager on August 24, 2025 UT. We confirm the previously reported CN and Ni outgassing and, for the first time, present the radial profiles of Ni and CN emission in 3I/ATLAS. We find characteristic $e$-folding radii of \Nitau~km for Ni and \CNtau~km for CN; this suggests that the Ni emission is more centrally concentrated in the comet's nucleus and favors hypotheses involving easily dissociated species, such as metal carbonyls or metal-polycyclic aromatic hydrocarbon molecules. Additional integral field spectroscopy after perihelion will provide an ongoing opportunity to determine the evolution of the radial distributions of species in interstellar comet 3I/ATLAS.  
\end{abstract}


\section{Introduction}\label{sec:intro}
Interstellar objects (comets and asteroids that pass through the solar system on hyperbolic orbits) offer a rare \citep[e.g.,][]{Do2018} window into the chemical and physical processes behind small bodies in other stellar systems \citep[e.g.,][]{Jewitt2023ARAA, Fitzsimmons2024}. Recently, the discovery of 3I/ATLAS \citep[][]{Denneau2025, Seligman2025, Tonry2025} by the Asteroid Terrestrial-impact Last Alert System (ATLAS; \citealp{Tonry2018a, Tonry2025}) has caused considerable excitement in the astronomical community. 3I/ATLAS is the third object of this type, following 1I/\Oumuamua\ \citep{Meech2017} and 2I/Borisov \citep{borisov_2I_cbet}. 

One avenue to understanding these objects is from cometary activity or lack thereof. This activity results from solar radiation heating the surface-layer materials \citep[e.g.,][]{Bessel1836, Whipple1950, Whipple1951, cowan1982}. When a comet warms, its volatile ices sublimate, lifting dust off the surface and creating the coma. Sunlight excites these gas molecules, which then re-emit light by resonance fluorescence—revealing the composition of the volatiles. For solar system comets, this provides a window into the primordial composition of the solar system \citep[e.g.,][]{Bodewits2024}. Interstellar objects enable similar studies for otherwise inaccessible stellar systems \citep[e.g.,][]{Jewitt2023ARAA}. 

The first two interstellar objects differed in their activity. No outgassing or coma was directly observed for 1I/\Oumuamua, despite extensive observations \citep[e.g.,][]{Meech2017, Ye2017, Jewitt2017, ISSI_1I_review, Trilling2018}. However, the only possible explanation for the non-gravitational acceleration of 1I/\Oumuamua was due to outgassing by CO or CO$_2$ \citep{Micheli2018}. On the other hand, 2I/Borisov had outgassing activity and a dusty coma \citep{Fitzsimmons:2019, Jewitt2019b, Cremonese2020, Guzik:2020, Hui2020, Kim2020, McKay2020, ye2020_borisov, yang2021}. The gas and dust of 2I/Borisov showed features in common with solar system comets, including CN \citep{Opitom:2019-borisov, Fitzsimmons:2019},$\mathrm{C}_2$ \citep{Lin2020}, [O~\textsc{I}] \citep{McKay2020}, OH \citep{Xing2020}, and $\mathrm{NH}_2$ \citep{Bannister2020} species, and $\mathrm{C}_2$, $\mathrm{C}_3$, and CH reported in post-perihelion observations \citep{Hoogendam26_KCWI}. 2I/Borisov had an unusually high abundance of CO \citep{Bodewits2020, Cordiner2020} and Ni outgassing \citep{Guzik2021, Opitom2021}. 

The recently discovered 3I/ATLAS is similar to 2I/Borisov: both showed cometary activity. \citet{Seligman2025} reported initial signs of activity, which were confirmed by subsequent studies \citep{Jewitt2025, Frincke25, Cordiner2025, Rahatgaonkar2025, Opitom2025, delaFuenteMarcos2025, Chandler2025, Lisse2025, Tonry2025}. The initial spectra exhibited red-sloped reflectance without strong emission features \citep{Seligman2025, Opitom2025, Puzia2025}. As it approached perihelion, outgassing increased. Absorption from large water ice grains in the coma \citep{Yang2025} and emission from CN \citep{Rahatgaonkar2025}, HCN \citep{Coulson25, Hinkle_JCMT}, Ni \citep{Rahatgaonkar2025, Hoogendam25_SNIFS, Hutsemekers25}, Fe \citep{Hutsemekers25}, $\mathrm{CO}_2$ \citep{Lisse2025, Cordiner2025}, CO \citep{Cordiner2025}, and a likely extended source of OH emission \citep{Xing2025} has been reported. 

Here, we present integral-field unit (IFU) observations of 3I/ATLAS from the Keck Cosmic Web Imager (KCWI; \citealp{Morrissey18}) on the Keck-II telescope. Integral field unit data provide several advantages over slit spectroscopy, including no slit loss of photons and the ability to undertake studies using spatial as well as spectral information. Previous IFU data for 3I/ATLAS includes a spectrum from the SuperNova Integral Field Spectrograph (SNIFS; \citealp{Lantz2004}) presented in \citet{Seligman2025}. This spectrum was obtained as part of the Spectroscopic Classification of Astronomical Transients survey, which typically observes transient phenomena \citep[e.g.,][]{Tucker2022, Tucker2026}. Further SNIFS observations are presented in a complementary study \citet{Hoogendam25_SNIFS}. A spectrum from the VLT-mounted Multi Unit Spectroscopic Explorer (MUSE, \citealp{Bacon10_MUSE}), taken shortly after discovery, also shows cometary activity \citep{Opitom2025}. Unfortunately, the wavelength coverage of MUSE is not blue enough to observe the CN and Ni features discussed in this work.

\section{Data}\label{sec:data}

\begin{figure}
\includegraphics[width=\linewidth]{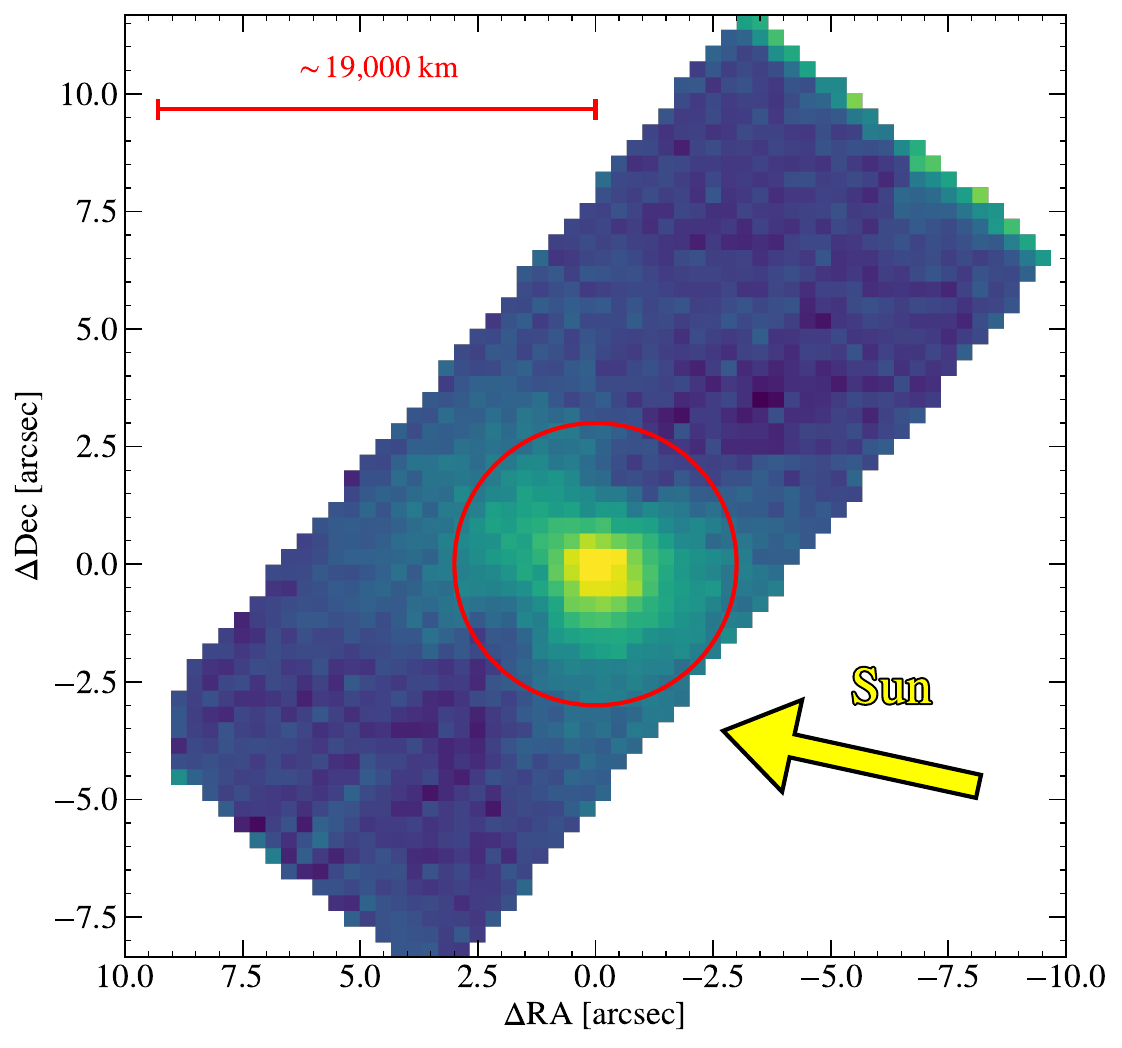}
\caption{3425 \AA\ to 5500 \AA\ whitelight image from the KCWI data. The strap through the center of the comet is an instrumental residual that \texttt{PypeIt} was unable to remove. The scale bar is the physical scale of the comet viewed from Earth. The aperture is the 3\arcsec\ aperture we used to extract our 1D spectrum. The yellow arrow denotes the direction to the Sun. The extended sun-to-target radius vector is 102.7 degrees.}
\label{fig:whitelight_broad}
\end{figure}

We obtained a KCWI spectrum on UTC 2025-08-24 06:06:49 (observation midpoint), along with two solar analogs and a flux calibration standard. 3I/ATLAS was at a heliocentric and geocentric distance of 2.75 and 2.60 au when our spectrum was taken. We used the small image slicer, providing an 8\arcsec\ by 20\arcsec\ field with a slice width of 0.35\arcsec. The blue channel used the BL grating, providing a spectral resolving power of $R\approx3600$. The blue central wavelength was 4500\AA. The reliable wavelength coverage in the blue is from 3400 \AA\ to 5500\AA. The total on-source integration time was 900~seconds. We reduced our KCWI data following standard \texttt{PypeIt} procedures \citep{pypeit:joss_pub}. We perform sky subtraction using the sky region from the standard star spectrum to prevent any self-subtraction of comet flux. We used calibration dome flats for flat-fielding and co-added the reduced, flux-calibrated cubes to create our final data cubes. \texttt{PypeIt} is not designed for non-sidereal objects, and the resulting flux calibration was disparate from the similar epoch ($\sim$1~day difference) spectra from both UH88/SNIFS (which is spectrophotometric) \citet{Hoogendam25_SNIFS} and the VLT/XSHOOTER data from \citep{Rahatgaonkar2025}. We correct our continuum flux to match these two other spectra (a factor of $\sim10$). We apply this correction to our spectrum and use the corrected flux values for subsequent analyses, since both instruments have well-calibrated reduction pipelines that have previously been used for non-sidereal objects. KCWI lacks a similarly well-tested pipeline for non-sidereal objects; this is the most accurate flux calibration we can obtain at this time. We extracted the spectrum using a 3\arcsec~aperture (beyond 3\arcsec, the aperture falls off the nearest chip edge). Figure \ref{fig:whitelight_broad} shows the 3400 \AA\ to 5500 \AA\ image from our KCWI spectrum. 

\section{Spectrospatial Analysis of 3I/ATLAS}\label{sec:analysis}

\begin{figure*}
\includegraphics[width=\textwidth]{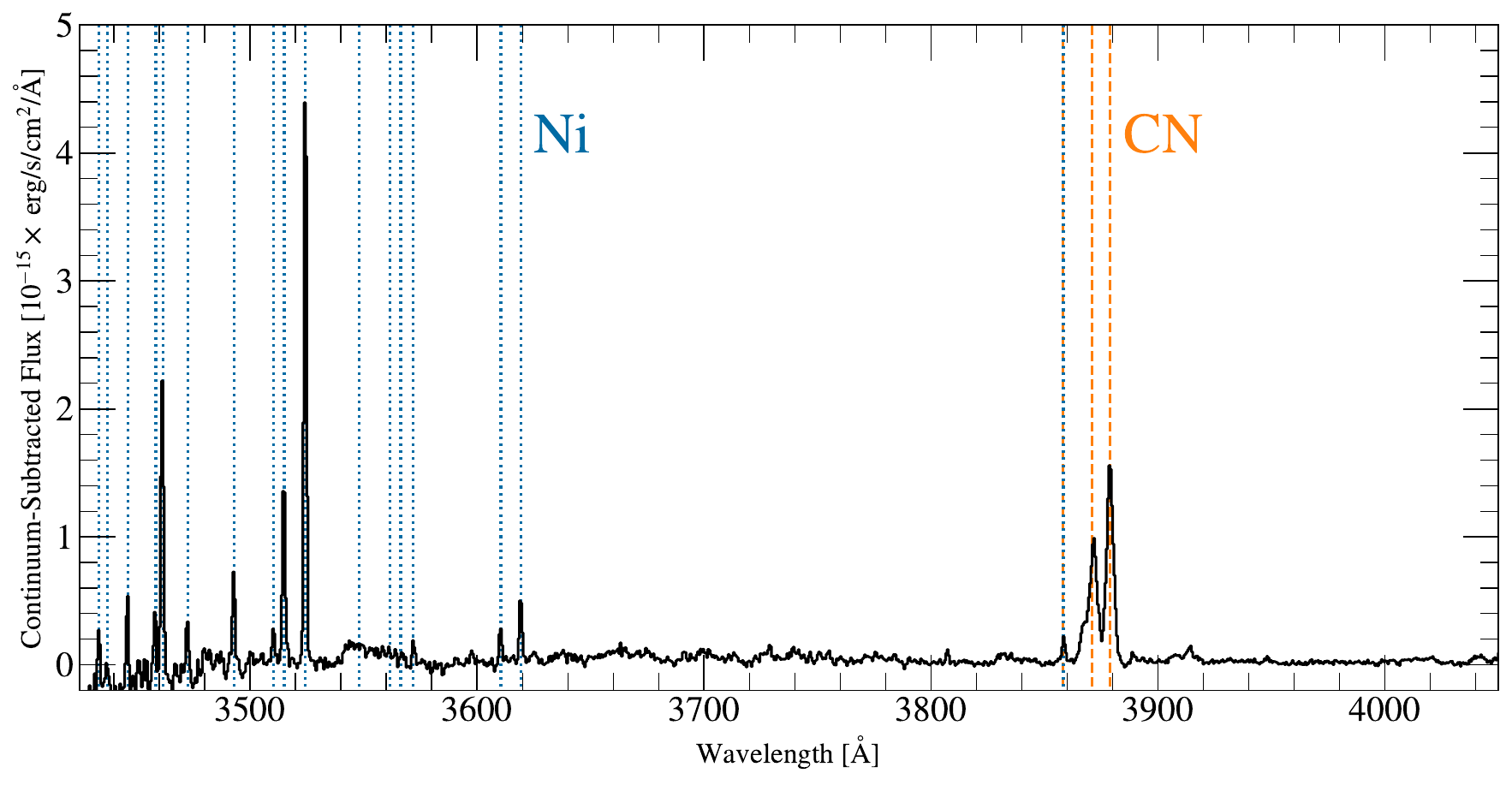}
\caption{The continuum-subtracted KCWI spectrum of 3I/ATLAS between 3425~\AA\ and 4050~\AA, extracted from a 3\arcsec\ aperture centered on the comet. The dotted blue and dashed orange lines denote Ni and CN, respectively. The emission feature at 3860~\AA\ overlaps with both potential CN and Ni features. Given that the Ni transition at that wavelength is not expected to be particularly strong, we attribute that emission to CN.}
\label{fig:spec1d}
\end{figure*}

To subtract the solar continuum contribution, we use HD~165290 as a solar analog. We model the continuum in the same manner as \citet{Rahatgaonkar2025}. The model is defined as 
\begin{equation}
    F_{\mathrm{cont}}(\lambda) = R(\lambda)\times F_{\odot\mathrm{analog}}\left[\lambda\left(1+\frac{v}{c}\right)+\delta\lambda\right],
\end{equation}

\noindent where ${R\left(\lambda\right)\equiv\frac{1}{S}\left(1+b_1\lambda+b_2\lambda^2\right)}$ is a second-order polynomial reflectance function normalized by a factor $S$ that compensates for flux differences between the comet spectrum and solar analog spectrum, and $F_{\odot\mathrm{analog}}$ is the solar analog spectrum shifted by fitted free parameters $v$ and $\delta\lambda$. $v$ and $\delta\lambda$ are nuisance parameters that account for the relative motion between Earth and the comet and Earth and the solar analog star, or, in other words, fit residuals from velocity and dispersion differences between the solar analog spectrum and the true solar spectrum. The parameters $b_1$ and $b_2$ quantify the reflectance slope. The solar analog has high SNR, so the continuum-subtraction contribution to the error budget is negligible.

Figure \ref{fig:spec1d} shows the 3425 \AA\ to \added{4000} \AA\ 1D continuum-subtracted spectrum from our KCWI datacube. Our spectrum extends to 5500 \AA, but lacks spectroscopic features redward of \added{4000} \AA\ (beyond CN). 

\subsection{Activity Signatures}
We observe previously reported CN and Ni features \citep[e.g.,][]{Rahatgaonkar2025}. \ion{Fe}{1} emission at $\sim$3722 \AA\ and $\sim$3728 \AA\ has previously been identified in Comet Ikeya-Seki (1965f) by two independent studies \citep{Preston1967, Slaughter1969} and other comets that also showed \ion{Ni}{1} emission \citep{Manfroid2021}, but we do not see evidence for this feature in 3I/ATLAS in our spectrum (a later detection is reported by \citealp{Hutsemekers25}). To place an upper limit on Fe emission in our spectrum, we insert a 3$\sigma$ Gaussian emission feature. We use the sigma-clipped RMS scatter of our continuum-subtracted spectrum as the 1$\sigma$ value. The flux upper limit of the \ion{Fe}{1} $\lambda$3720 feature is {$F_\mathrm{Fe}\,=\,9.9\times\,10^{-16}$\,erg\,s$^{-1}$\,cm$^{-2}$\,\AA$^{-1}$}.

A simple \citep{Haser:1957} model was used to convert the measured CN line flux into a gas production rate. The number of photons emitted per molecule per second (the so-called g-factor) and the scale lengths were taken from \citep{A'Hearn:1995}. We assumed that the gas escapes isotropically from the nucleus at a constant velocity, and adopted a mean expansion speed of $0.8 \times r^{-0.6}$~km~s$^{-1}$, following \citep{Biver:1999}, where $r$ is the heliocentric distance in au. The derived CN production rate is 
$\mathrm{Q(CN)}=(3.9\pm0.5)\times10^{24}$~molecules~s$^{-1}$.

\subsection{Radial Distribution of CN and Ni}\label{sec:radial}

\begin{figure*}
\includegraphics[width=\textwidth]{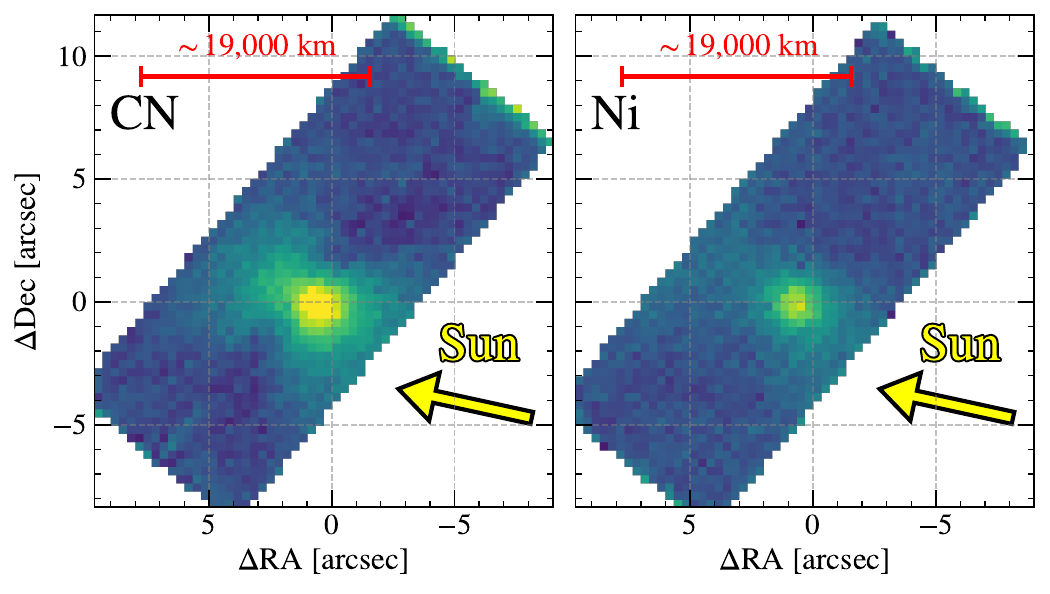}
\caption{Comparison of narrow-band images from the KCWI data cube for CN (\emph{left}, from 3865\AA\ to 3885\AA) and Ni (\emph{right}, from 3605 \AA\ to 3625 \AA). Both plots have a normalized color scale to account for differences in flux between the two lines. The scale bar is 10\arcsec.}
\label{fig:KCWI_whitelight_comp}
\end{figure*}

An advantage of our IFU data is the immediate availability of spectro-spatial information about 3I/ATLAS. Slit-based techniques for studying radial distributions for comets involve dithering the slit between a central position and an offset position. Comet asymmetries in the coma may bias results from this technique, which has a limited field of view. Our data extend over 3\arcsec, or $\sim$5700~km, before the coma reaches the nearest chip edge. We leverage this unique data to analyze the spatial distribution of CN and Ni in 3I/ATLAS. 

Figure \ref{fig:KCWI_whitelight_comp} shows the KCWI 2D narrow-band images of two similar-strength CN (\emph{left}, from 3865\AA\ to 3885\AA) and Ni (\emph{right}, from 3605 \AA\ to 3625 \AA) features. An extended coma is visible for CN, whereas the Ni emission has a more central profile. 
Figure \ref{fig:KCWI_whitelight_comp} also shows anisotropy in the distribution of CN, with higher flux concentrations visible in both the sunward direction---where a dust plume has been observed in most observations of 3I/ATLAS since its discovery \citep{Seligman2025, Chandler2025, Jewitt2025, Cordiner2025}---and the anti-sunward direction, where anti-velocity directed cometary tails are spread out by a combination of \added{radiation pressure and solar wind gas drag}. 3I/ATLAS has indeed developed an anti-solar tail as of late August.

\begin{figure*}
\includegraphics[width=0.95\linewidth]{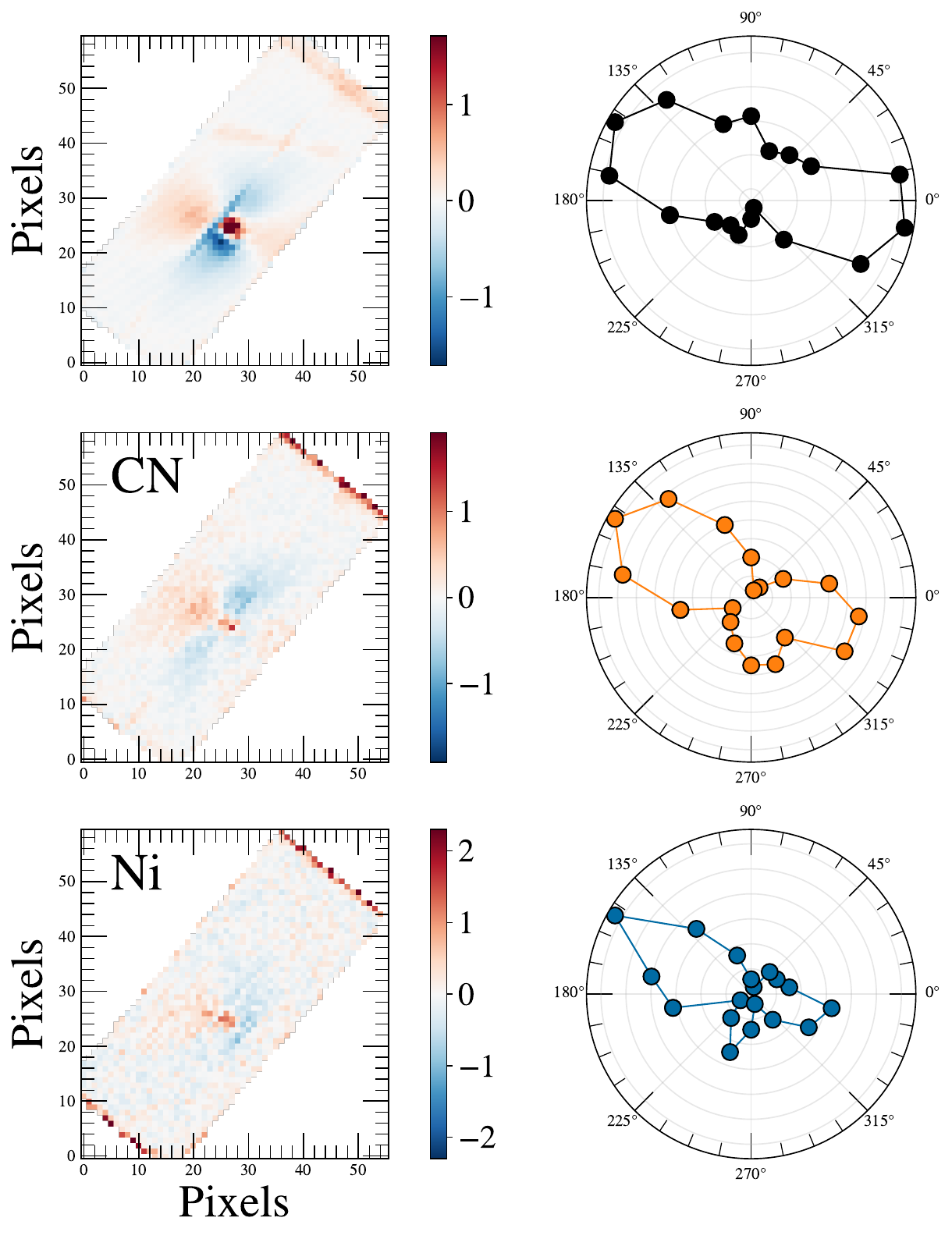}
\caption{Determination of jets in the radial profiles of 3I/ATLAS.
         \emph{Left:} Model-subtracted data for white light (top), CN (middle), and Ni (bottom) regions. Strong solar and anti-solar flux excesses are visible in the broadband images. The flux values are rescaled to be near unity. 
         \emph{Right:} Model-subtracted flux excess as a function of azimuth. The radial coordinate values are re-scaled for each image; the extent of Ni emission is much more central than the broadband emission or CN, and the flux excess extends further for the broadband than for the CN and Ni. 
         }
\label{fig:angular-profile}
\end{figure*}

The angular flux profile of 3I/ATLAS is shown in Figure \ref{fig:angular-profile}. The \added{left column shows the profile-subtracted data. The subtracted image reveals flux overdensities in the solar and anti-solar directions. The perpendicular flux underdensities likely result from symmetric profile oversubtraction in those regions. The right column shows the azimuthal profiles calculated from the residual image maps.}

\begin{figure}
\includegraphics[width=\linewidth]{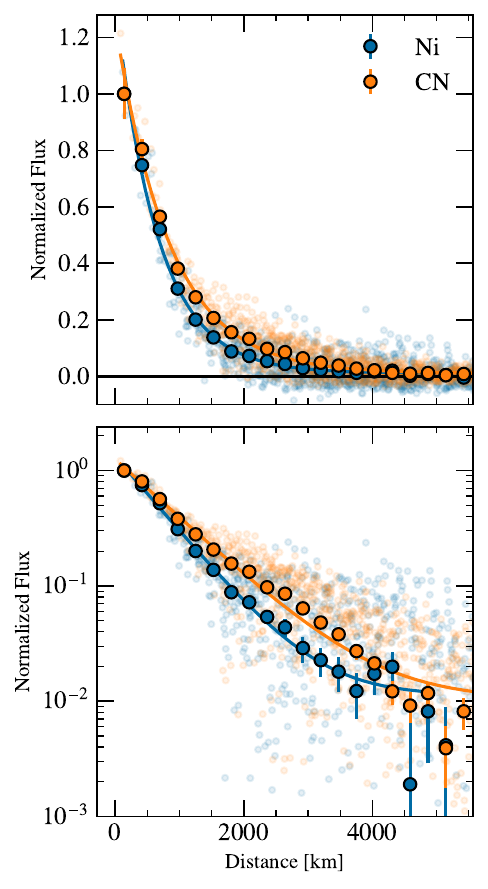}
\caption{Measured radial differences between CN and Ni using the same ranges as Figure \ref{fig:KCWI_whitelight_comp} in linear (top) and log (bottom) spacing. The maximum flux values for each feature are normalized to 1 to correct for (potential) differences in continuum luminosity. The sky background away from the comet is subtracted. Small points are the raw data, with larger, less transparent points showing the binned data. The lines are exponential functions fit to the data.} 
\label{fig:KCWI_profiles}
\end{figure}

The azimuthally averaged flux as a function of physical radial distance is shown in Figure \ref{fig:KCWI_profiles}. We bin the radial profiles into one-pixel bins that correspond to $\sim$278~km given the pixel scale (0.147\arcsec/pixel) and cometary distance. We normalize the flux to the maximum binned flux value. The Ni is more centrally concentrated than CN, with the majority of Ni flux coming from the innermost 2000 km of 3I/ATLAS. On the other hand, the radial profiles suggest that CN, though still concentrated near the nucleus, extends farther out into the coma.

We fit the radial profiles of CN and Ni with an exponential decay model defined as 
\begin{equation}
    A\times\exp\left[\frac{-x}{\tau}\right] + C    
\end{equation}
where $\tau$ is the characteristic $e$-folding length scale of the radial profile (and $A$ and $C$ are nuisance parameters). To account for differences in continuum level, we normalize the data to the maximum flux value. The fit is iterated 10 times; after each fit, outliers are sigma-clipped, and the sigma-clipped data are refit. 
For Ni, we measure \Nitau~km (with $A=1.233\pm0.034$ and $C=0.006\pm0.002$), and for CN \CNtau~km (with $A=1.028\pm0.018$ and $C=0.010\pm0.002$).

\section{Potential Physical Drivers of Ni Emission in Interstellar Comets}

The recent work by \citet{Rahatgaonkar2025} discusses three possibilities for the Ni emission: metal carbonyls, such as $\mathrm{Ni(CO)}_4$ and $\mathrm{Fe(CO)}_5$, are present in the comet \citep[e.g.,][]{Manfroid2021, Guzik2021}; metal-polycyclic aromatic hydrocarbons \citep[PAH,][]{Tielens08} that photofragment and release Ni and PAHs; and a hybrid scenario \citep[e.g.,][]{Brownlee06, Bardyn17} that invokes ``in-situ'' formation of Ni-carbonyls with Ni released from Ni-sulfides. The radial distribution of Ni is a critical diagnostic to evaluate each scenario.

In the metal carbonyl scenario, $\mathrm{Ni(CO)}_4$ is more stable than its ferrous counterpart $\mathrm{Fe(CO)}_5$ in the presence of water and oxidants. This enables trace $\mathrm{Ni(CO)}_4$ to persist where $\mathrm{Fe(CO)}_5$ would not, giving the observed \ion{Ni}{1} emission without any \ion{Fe}{1} counterpart. Additionally, $\mathrm{Ni(CO)}_4$ quickly dissociates, which would result in a central concentration compared to other species, such as CN. 

Metals such as Ni may attach to PAHs \citep{Tielens08} to form Ni+PAH molecules (e.g., Ni-naphthalene $\mathrm{Ni}(\mathrm{C}_{10}\mathrm{H}_{8})$ potentially). These molecules are easily unbound by absorbing light and may produce centrally concentrated Ni, as observed in 3I/ATLAS.

The hybrid scenario invokes UV sputtering or thermochemical erosion to produce Ni atoms from Ni-sulfides such as $(\mathrm{Fe,Ni})_9\mathrm{S}_8$. If the macroenvironment is CO-rich, then these sulfides carbonylate to $\mathrm{Ni(CO)}_4$. The $\mathrm{Ni(CO)}_4$ subsequently undergoes a similar photodissociation as the metal-carbonyl model described above. This ``in-situ'' formation of $\mathrm{Ni(CO)}_4$ predicts that Ni should be strongly concentrated near the nucleus, with potential anisotropies where dust jets intersect outflows of CO \citep[e.g.,][]{Guzik2021}. We observe a central Ni concentration; however, Figure \ref{fig:KCWI_whitelight_comp} shows symmetric Ni emission. While \citet{Cordiner2025} reports a CO detection from \emph{James Webb Space Telescope} data, compared to 2I/Borisov and other comets, 3I/ATLAS is not CO-rich, and its distant activity was mostly driven by $\mathrm{CO}_2$. 

An emerging trend is that both 2I/Borisov and 3I/ATLAS appear to show Ni/Fe ratios significantly above the solar value, as do the solar system comets in whose comas both metals have been measured at heliocentric distances too great for bulk refractory materials to sublime \citep{Manfroid2021, Opitom2021}. As discussed above, Fe-related carbonyls and PAH-based molecules may face greater obstacles to either formation, survival, or both, making them less common. 

Another anomalous abundance is that of Fe/Ni relative to $\mathrm{H}_2\mathrm{O}$. 2I/Borisov \citep{Opitom2021} and 3I/ATLAS are similar, based on the late-August ($r_h\approx2.85$ au) Q(Ni) from \citet{Rahatgaonkar2025} and the Q($\mathrm{H}_2\mathrm{O}$) from \citep[][]{Xing2025}. While water ice can sublimate at this heliocentric distance, the activity in 3I/ATLAS at 2.85 au appears to be driven primarily by $\mathrm{CO}_2$ rather than $\mathrm{H}_2\mathrm{O}$, as the sublimation flux of water ice at this distance may be insufficient to lift dust from the nucleus surface efficiently. Despite highly abundant CO, 2I/Borisov had a Q(Fe+Ni)/Q(CO) ratio that was similar to solar system comets \citep{Opitom2021}. Furthermore, comparing the \citet{Rahatgaonkar2025} Q(Ni) measured on Aug 9 (or Jul 31) with the Q(CO) measurement on Aug 6 from \citet{Cordiner2025}, Q(Ni)/Q(CO) looks in line with the solar system trend (similar to 2I/Borisov), but Q(Ni)/Q($\mathrm{H}_2\mathrm{O}$) looks $\sim$2 orders of magnitude high, again similar to 2I/Borisov \citep{Opitom2021}.

The Q(Ni)/Q(CN) ratio from \citet{Rahatgaonkar2025} is $\sim$0.1, which is higher than 2I/Borisov and orders of magnitude above the solar system comet median \citep{Bromley21}. This type of analysis, while only briefly discussed here, could help narrow down the source(s) of Fe and Ni in both solar system and interstellar comets and evaluate hypotheses for why these interstellar comets (so far) have different compositions (water-poorer relative to CO and $\mathrm{CO}_2$) compared to a vast majority of solar system comets.

\section{Conclusions}
We present an IFU spectral analysis of the third interstellar comet 3I/ATLAS using data from Keck-II/KCWI when 3I/ATLAS had a heliocentric distance of 2.75 au. We confirm previously reported CN and Ni activity, and measure the radial profiles of both CN and Ni, finding that the Ni emission is more centrally concentrated than the CN. Ni has a characteristic $e$-folding radius $\sim$200~km shorter than CN. This further supports a growing picture that Ni emission in comets arises from short-lived parent species, whose short lifetime is imprinted on the radial distribution of Ni. We also find evidence for asymmetric profiles for the broadband and CN and Ni narrowband images, with brighter integrated fluxes found to align with the solar and antisolar direction. This may be evidence for previously reported dust plumes in 3I/ATLAS \citep{Seligman2025, Chandler2025, Jewitt2025, Cordiner2025}.

The presence of gaseous Ni in solar system comets and in 2I/Borisov and 3I/ATLAS suggests similarities between the unknown birthplaces or interstellar processing of the two latest interstellar interlopers and our own solar system. The lack of Fe emission in 3I/ATLAS differs from the solar system comets presented in \citet{Manfroid2021, Guzik2021} and 2I/Borisov \citep{Opitom2021}, but recent detections may indicate that Fe emission is triggered at closer heliocentric distances \citep{Hutsemekers25}. This could provide new insights into how the chemistry, metallicity, and radiative evolution of stellar disks influence the formation and evolution of their small bodies. 

The driving mechanism for Ni emission in the solar system and interstellar comets may differ, but independent of the physical driver, the measurement of Ni in a growing number of interstellar comets offers a promising pathway to trace the parent chemistry and metallicity of extrasolar planetary systems. Future IFU data will also provide spectrospatial information about the Fe radial distribution, enabling further comparative analysis between Fe, Ni, and other volatiles. The upcoming Legacy Survey of Space and Time \citep{LSST_2019} will discover and observe more interstellar comets, especially at greater heliocentric distances. As the number of these objects increases, population-level studies of the metal content in these objects and its dependence on heliocentric distance will be feasible, providing new insights into the extrasolar systems in which they form. 

\begin{acknowledgments}
% We sincerely thank the anonymous referee for a helpful review that improved the manuscript.

We thank Ryan Cooke for helpful discussions about KCWI reductions using PypeIt. 
We thank Thomas Puzia, Jaun Pablo Carvajal, John Noonan, and Cyrielle Opitom for helpful discussions about 3I/ATLAS and this work. 

This work benefited significantly from the expertise and assistance of Keck Support Astronomer Greg Doppmann, who expedited access to the observing data and calibration files, and OA Julie Renaud-Kim, who was instrumental in getting fast setups on non-sidereal targets working, enabling the efficient use of telescope time.

W.B.H. acknowledges support from the National Science Foundation Graduate Research Fellowship Program under Grant No. 2236415. 

K.J.M., J.J.W., and A.H.\ acknowledge support from the Simons Foundation through SFI-PD-Pivot Mentor-00009672. B.J.S, K.J.M, and W.B.H. acknowledge support from NSF (AST-2407205) and NASA (HST-GO-17087, 80NSSC24K0521, 80NSSC24K0490, 80NSSC23K1431) grants. 

J.T.H. acknowledges support from NASA through the NASA Hubble Fellowship grant HST-HF2-51577.001-A, awarded by STScI. STScI is operated by the Association of Universities for Research in Astronomy, Incorporated, under NASA contract NAS5-26555.

This research made use of \texttt{PypeIt}\footnote{\url{https://pypeit.readthedocs.io/en/latest/}},
a Python package for semi-automated reduction of astronomical data \citep{pypeit:joss_pub, pypeit:zenodo}.

\end{acknowledgments}

\bibliography{sample701}{}
\bibliographystyle{aasjournal}

\end{CJK*}
\end{document}